\documentclass[traditabstract]{aa}
\usepackage{graphicx}
\usepackage{amssymb}
\usepackage{txfonts}
\usepackage{deluxetable}
\def\Mj{\,$\mathrm{M}_\mathrm{J}$}
\def\mc2{\multicolumn{2}{c}}

\begin{document}

\title{The Minimum Jeans Mass, Brown Dwarf Companion IMF, and 
Predictions for Detection of Y-type Dwarfs}

\author{B.~Zuckerman\inst{1} \and
Inseok Song\inst{2}}

\institute{Dept. of Physics \& Astronomy and Center for Astrobiology,
	   University of California, Los Angeles,
	   475 Portola Plaza,
	   Los Angeles, CA 90095--1547, USA\\
	   \email{ben@astro.ucla.edu}
\and     Department of Physics \& Astronomy,
	   The University of Georgia,
	   Athens, GA 30605, USA\\
	   \email{song@physast.uga.edu}
  }

\date{Received; accepted}
\authorrunning{Zuckerman \& Song}
\titlerunning{BD companion IMF \& Y-dwarfs}

\abstract {
Cool L- and T-type objects were discovered first as companions to stars in 1988
and 1995, respectively.  A certain example of the yet cooler Y-type spectral
class ($T_{eff}$$\lesssim$500\,K?) has not been seen. Recent infrared imaging
observations of stars and brown dwarfs indicate that substellar companions with
large semi-major axes and with masses less than the brown dwarf/giant planet
dividing line ($\sim$13.5\,\Mj) are rare.  Theoretical considerations of Jeans
mass fragmentation of molecular clouds are consistent with this minimum mass
cutoff and also with the semi-major axis (hundreds of AU) characteristic of the
lowest mass imaged companions. As a consequence, Y-class companions with large
semi-major axes should be scarce around stars $<2$\,Gyr old, and also around
substellar primaries of all ages. By focusing on brown dwarf companions to
young stellar primaries, it is possible to derive a first estimate of the brown
dwarf IMF over the entire range of brown dwarf masses (13\,\Mj\ to 79\,\Mj) --
the number of companion brown dwarfs is proportional to mass to the
$-1.2\pm0.2$ power.  }


\maketitle

\section{Introduction}
The temperature of the coolest measured substellar dwarf objects has diminished
from 1900\,K in 1988 (GD\,165B, Becklin \& Zuckerman 1988; Kirkpatrick et al.
1999), to 960\,K in 1995 (Gl\,229B, Nakajima et al. 1995; Geballe et al. 2002),
to 800\,K in 2000 (Gl\,570D, Burgasser et al. 2000; Geballe et al. 2001), to
650\,K in 2007 (ULAS\,J0034-00, Warren et al 2007). Recently, Delorme et al.
(2008) identified a 620\,K field brown dwarf that, along with ULAS\,J0034-00,
show a suggestion of an ammonia absorption feature in the H-band. They note
that if the apparent ammonia feature deepens at lower effective temperatures,
then $\sim$600\,K will be a natural break point between the T and Y spectral
types.  In parallel, imaging searches for objects of planetary mass have
revealed 2M1207b with a mass $\sim$5 times that of Jupiter (Chauvin et al.
2004, 2005a; Song et al. 2006) or perhaps $\sim$8\,\Mj\ (Mohanty et al. 2007),
along with a handful of substellar companions (Table~1) at or just above the
planet/brown dwarf boundary (13.5\Mj) as defined by the IAU.  Published
(Masciadri et al. 2005; Luhman et al. 2007a; Kasper et al.  2007a; Lafrenie\'re
et al. 2007) and as yet unpublished (G.  Chauvin et al., in preparation; J.
Farihi et al., in preparation; C. Marois et al, in preparation) imaging
searches with adaptive optics (AO) systems on the VLT and on Keck, with the
NICMOS camera on HST, and with IRAC on Spitzer, are sensitive to objects with
temperatures $<800$\,K as well as companion masses as small as a few Jupiters
in the case of many target stars.  

To set limits on the masses of planets that can be detected at a given 
separation from a given target star, the standard procedure in most 
planet-imaging survey papers is to employ a series of Monte Carlo simulations 
of an ensemble of extrasolar planets around each star.  The luminosity of a 
planet, based on theoretical mass-luminosity-age calculations, at each 
semimajor axis is compared with the measured minimum detectable brightness in 
each annulus around a target star.  In this way it is possible to determine 
just how close to any given star a planet of any given mass might be 
observable.  For example, in a sample of 85 nearby young 
stars, \cite{Lafreniere} were sensitive to planets more massive than 2 
\Mj\ with a projected separation in the range 40$-$200 AU around a typical
target.  They found none, and at a 95\% confidence level concluded that at 
most 
12\% of stars harbor a planet more massive than 2 \Mj\ between 50 and 295 
AU.
\cite{Nielsen} found no planets orbiting 60 stars and concluded with 95\% 
confidence that the fraction of stars with planets with semimajor axis between 
20 and 100 AU and mass above 4 \Mj\ is $<$20\%.   \cite{Kasper07a} derived a 
frequency of giant planets with masses above 2$-$3 \Mj\ at separations larger 
than 30 AU around nearby G, K, and M-type stars to be $\leq$5\%.

Notwithstanding the above published and unpublished survey
sensitivities down
to planets of a few Jupiter masses,
Table~1 includes all companions reported to date with masses probably less than 
20 times that of Jupiter. In the present paper we gather together results from 
the many papers listed above and from some additional papers to: (1) 
Demonstrate that substellar companions at large semi-major axes (beyond those 
accessible to the techniques of precision radial velocities and microlensing) 
can be accounted for in a Jeans-mass fragmentation model. (2) Derive a 
first 
estimate of the brown dwarf companion initial mass function (IMF) over the 
entire range of brown dwarf masses.  (3) Infer that Y dwarfs may be detected 
occasionally as companions to old stars, but will be particularly rare as 
companions to stars with ages $<$2 Gyr.

We first consider the significance of Table~1 in the
context of formation scenarios for brown dwarfs and massive
planets with semi-major axes sufficiently large for imaging
detection with existing instruments.  Then we discuss implications
of Table~1 for imaging discovery of Y-type substellar companions
(Kirkpatrick 2000).  This new spectral class, cooler than T-type,
may begin to appear at effective temperatures around 500\,K (e.g.,
Burrows et al. 2003).

\begin{table*}
\begin{minipage}[t]{\textwidth}
\begin{center}
\caption{Lowest mass companions (M $\leq$20 \Mj) imaged to date.}
\begin{tabular}{cccccccl}
\hline \hline
Object & \mc2{Sp. Type} & Age & $M_{pri}$ & $M_{sec}$ & Sep.  & Ref. \\
\cline{2-3}
& Primary & Secondary & (Myr) & (M$_{\odot}$) & (\Mj) & (AU) & \\
\hline
2M1207     & M8    & L5    &$    8     $&  0.025 &  5$^a$  &  46  & Chauvin et al (2004)                 \\
AB Pic     & K2V   & L1    &$   30     $&  0.84  & 14  & 248  & Chauvin et al (2005c)      \\
Oph 11     & M9    & M9.5  &$    5     $&  0.0175& 15  & 237  & Close et al (2007), Luhman et al (2007b)   \\
GQ Lup     & K7V   & L1.5  &$    3?    $&  0.7   & 17  & 100  & Neuhauser et al. (2005)         \\
HN Peg     & G0V   & T2.5  &$   200^b  $&  1.0   & 18  & 795  & Luhman et al.  (2007a)            \\
TWA 5      & M1.5  & M8/8.5&$    8     $&  0.40  & 20  &  98  & Lowrance et al. (1999)  \\
LP 261-75  & dM4.5e& L6    &$ 100-200  $&  0.12  & 20  & 506  & Kirkpatrick et al. (2000)       \\
\hline
\end{tabular}
\end{center}
{\footnotesize
The mass listed for GQ Lup B is the average of the
geometric means of the range of likely masses given in McElwain et
al (2007) and Marois et al (2007). \\
$^a$ Mohanty et al (2007)
argue that the mass of 2M1207b is 8$\pm$2 M$_J$ \\ 
$^b$ We list the age of the HN Peg system as 200\,Myr while
\cite{HNPeg} estimated an age of 300\,Myr.  We prefer the younger age
for the following reasons.  The lithium 6708\,\AA\  equivalent width
is $\sim$105\,m\AA\ based on an average of independent measurements of
101 and 110\,m\AA\ by \cite{Gaidos}  and \cite{Chen}.   With a
Johnson $B-V$ = 0.59\,mag, from Hipparcos, or derived from Tycho-2
data (Bessell 2000), Figure~3 in \cite{ARAA} and Figure~1 in
\cite{HR3070} indicate a lithium age for HN~Peg slightly older than
that of the Pleiades.  The Galactic space motion $UVW$ ($-15, -22,
-11$; Nordstro\"m et al. 2004) is also consistent with those of many young
(Pleiades age or younger) stars in the solar vicinity.  The
logarithmic ratio of X-ray luminosity (as measured in the ROSAT All
Sky Survey) to bolometric luminosity is $-4.43$.   This is similar to
the Pleiades or the slightly older Carina-Near moving group (see Fig.
2 in Zuckerman et al. 2006).  \cite{HNPeg} mention chromospheric activity
suggestive of an age of 0.35\,Gyr.  But for a rotation period of 4.91
days (for HN~Peg, Gaidos et al. 2000), \cite{CaAge} show that, for
youthful stars, CaII emission consistently overestimates stellar ages
by a factor of at least a few relative to Li/X-ray/UVW ages.  This
rotation period and B-V can be used to derive a gyrochronology age of
247$\pm$42~Myr (Barnes 2007).  Combination of this age with the
somewhat younger ages indicated by lithium, UVW, and X-ray flux,
suggests 200\,Myr as the most probable age of HN Peg.
}
\end{minipage}
\end{table*}

\section{Discussion}

\subsection{Fragmentation by gravitational instability}

Thirty years ago, \cite{LL} published a paper ``The minimum Jeans mass, 
or when fragmentation must stop'' in which they derived the minimum 
fragment mass in a typical dark molecular cloud.  This mass, about 7 
times that of Jupiter, was found to be insensitive to properties of the 
interstellar dust and cosmic ray heating flux, while various processes 
(e.g., rotation, magnetic fields, late accretion) would normally be 
expected to increase this minimum mass.  Furthermore, as Low \& 
Lynden-Bell pointed out, (a) for a fragment to split it must have at 
least twice the minimum mass and (b) since a Jeans mass perturbation 
has a zero growth rate, one would expect a real growing perturbation to 
be somewhat more massive than a Jeans mass.  Subsequently, Bate and 
collaborators (see Bate 2005 and references therein), 
\cite{BW05}, \cite{PN04}, \cite{Padoan05}, \cite{Whitworth}, and 
Padoan et al (2007) considered additional processes such as turbulence, 
shock compression and the role of magnetic fields and derived fragment 
masses that could be as low as three times Jupiter's mass.

As indicated in Table~1 and discussed below, at present, no
imaged companion is known with a mass clearly below 7 Jupiter masses
despite the fact that the imaging searches listed in Section~1 should
have been sensitive enough to detect some at wide separation if they
are copious. Indeed the apparent pile-up of minimum substellar masses
at about twice this value (Table~1) is qualitatively consistent with
the considerations of Low \& Lynden-Bell outlined in the previous
paragraph.  

Notwithstanding that AO and HST imaging programs focus on
detection of massive planets at the smallest measurable
separations, typically $1''-2''$ or $<50$\,AU, the characteristic
separation in the Table~1 binaries is hundreds of AU.  Depending
on projection effects, even the semi-major axis of 2M1207b might be
much larger than 46\,AU.   Low \& Lynden-Bell derive an average
distance between fragments at last fragmentation of a few 100 
AU, consistent with the typical separation of pairs listed in Table~1.
\cite{Rafikov} considered the possibility of giant planet
production by gravitational instability in protoplanetary disks.
His model can produce massive planets at $\sim$100\,AU with masses
similar to those listed in Table 1, if the initial disk
mass is at least a few tenths of a solar mass.\footnote {\cite{Rafikov} 
presents arguments against gravitational instability
occurring at 1 and at 10\,AU, semi-major axes appropriate to the
planets discovered by the precision radial velocity (PRV)
technique.  However, as noted by \cite{ABPic} and by \cite{Rice}
inspection of the PRV database indicates that the well-known
correlation between high stellar metallicity and the existence of
planets may not obtain for stars with the highest mass planets  
($>7$\Mj).  That is, the relatively few highest mass PRV planets   
may have formed by gravitational collapse, which, compared to core
accretion, is relatively insensitive to metallicity.} Further 
consideration of gravitational fragmentation in massive young
proto-planetary disks can be found in Stamatellos \& Whitworth (2008).

Objects of planetary mass might exist at large semi-major axes as
a consequence of physical mechanisms other than fragmentation.
For example, it has been suggested that three body interactions
involving a star and either two nearby orbiting planets or a
second star with a planet might gravitationally eject a planet to a
large semi-major axis.  However, between the mass of Jupiter and
14 times this mass, the number of planets discovered by the
precision radial velocity technique rises steeply with decreasing
mass (Marcy et al. 2005; Lovis et al. 2006).  Thus, given that the AO and HST
programs listed in Section 1 are often sensitive at large
separations to planets down to a few Jupiter masses, and that in
3-body interactions one expects the lowest mass object to be
ejected, the distribution of masses given in Table 1 is
inconsistent with the ejection model.

\subsection{Y type companions}

The upper temperature limit for a Y-type object is not known; in the 
following we assume it to be 500\,K.  Based on Table 1 and on the above 
considerations, the percentage of stars with companions with large 
semi-major axes and mass $\lesssim15$ Jupiter masses appears to be very 
small indeed.  According to \cite{Baraffe} and \cite{Burrows03} the 
time needed for a 15 Jupiter mass object to cool to 500\,K is 
$\sim$2\,Gyr.  Given the many hundreds of young stars with ages 
$\lesssim100$\,Myr that have been searched with HST and ground-based AO 
down to this mass, that only three such systems are now known (AB~Pic, 
HN~Peg and GQ~Lup) indicates that even at 2\,Gyr, Y-dwarfs should be 
rare as companions to stars in wide orbits. In other words, for 
Y-type companions 
to be abundant at $\sim$2\,Gyr, many early to mid-L companions 
($\lesssim15$\Mj) to young ($\lesssim100$\,Myr) stars should have been 
detected in imaging searches of the sort listed in Section 1. But such
L-type companions are very rare.

The situation regarding Y-type secondaries of $\sim$15 Jupiter
mass where the primary is a brown dwarf is probably even more 
unfavorable. Although two substellar binaries belonging to very young 
associations (2M1207 and Oph 11) appear in
Table 1, no comparable systems are known to exist among the field
brown dwarfs, not even those as young as AB~Pic and HN~Peg.  Indeed, a
recent search for wide companions to 132 M7--L8 primaries in
the field came up completely empty handed (Allen et al. 2007).  The low
binding energies as displayed in Figure 1 may be a clue as to why
older analogs to 2M1207 and Oph 11 are so rare.  However, this
connection need not necessarily be straightforward because
\cite{Burgasser03} argue that the deficiency of field brown dwarf
binaries with semimajor axes $>10$\,AU cannot be explained as due
to disruption over Gyrs by encounters with stars and giant
molecular clouds.  Additional discussion of this point and others
may be found in \cite{Burgasser03, Burgasser06} and in
\cite{Allen} and \cite{Close}.   If the separation of a minimum mass 
fragment ($\lesssim15$\,\Mj) is as large as a few 100\,AU as described
above, then such fragments will not be found among the population
of 10\,AU brown dwarf binaries previously discovered with HST and with 
ground-based AO.

\begin{figure}
\begin{center}
\includegraphics[clip,width=0.95\columnwidth]{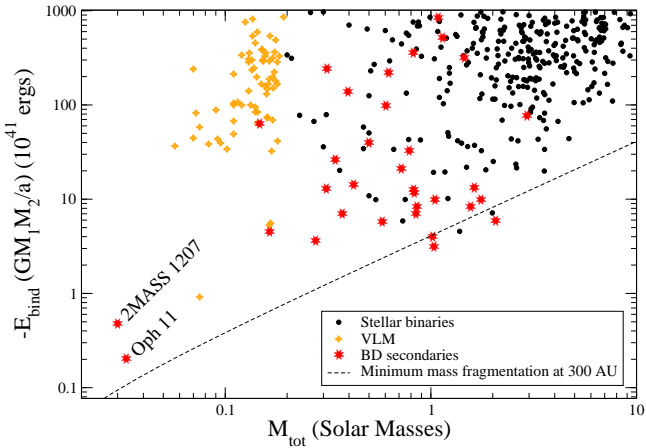}
\end{center}
\caption{Binding energies of lowest mass companions known to date.
For 2M1207 and Oph 11 (only) both the primary and secondary are
substellar.  Binaries with stellar primaries are from
\cite{Fisher} and \cite{Tokovinin}.  Very low mass (VLM) binary
data are from N. Siegler's VLM archive
http://paperclip.as.arizona.edu/\~nsiegler/VLM\_binaries/ }
\end{figure}

\subsection{IMF for brown dwarf secondaries to stellar primaries}

In Tables 1 and 2 we have gathered from the literature as many brown 
dwarf secondaries to stellar primaries as we could find, and then 
plotted their number distribution with mass (M) in Figure 2.  Table 1 
contains two systems with brown dwarf primaries, these do not appear in 
Figure 2.  In comparison with the steep Salpeter 
distribution for intermediate mass stars -- number, N, proportional to 
(M$^{-2.35}$) -- the Fig. 2 distribution appears much flatter.  This 
figure displays a striking separation in companion mass between 
youthful and oldish systems.  It seems clear that the systems used to 
detect brown dwarfs around oldish stars have generally been 
insufficiently sensitive to reveal cool, low mass, brown dwarfs.  
Therefore, we use only the dozen systems with ages $\lesssim300$ Myr to 
derive the brown dwarf companion IMF; that is, N is proportional to 
mass as M$^{-1.2\pm0.2}$ between 13 and 79\,\Mj.

For free-floating low mass objects, Allen et al (2005) derived N
proportional to (M$^{-0.3\pm0.6}$), which covered masses only down to
about 40\,\Mj.  Allen's sample is the nearby stars. A recent paper by
Anderson et al (2008) focuses instead on the IMF of free-floating low mass
stars and brown dwarfs in six very young clusters (1$-$2 Myr old) and in
the Pleiades at distances between 125 and 830 pc from Earth.  They probe
masses down to about 30 M$_J$ and deduce that the mass function is falling
as one passes from the stellar to the brown dwarf regime (i.e., in the
expression for (dN/dM), the exponent on M is positive, rather than
negative as found by Allen et al and in the present paper).  
With their large error bar ($\pm$ 0.6), the Allen et al (2005) result is
not incompatible with the falling IMF deduced by Anderson et al. (2008).

However, we are troubled by a number of aspects of the Anderson et al 
(2008) conclusion.  Based on results for the Taurus star-forming region 
reported by Konopacky et al (2007), the number of unresolved binaries is 
apt to be substantially larger than assumed by Anderson et al (see their 
Section 4).  The Konopacky paper is not cited by Anderson et al.  In 
addition, the Anderson et al analysis implicitly assumes that low mass 
objects (i.e., brown dwarfs) are fully formed and would be noted at their 
final masses by a cluster age of 1 or 2 Myr.  This assumption may not be 
valid.

Both Allen et al (2005) and Anderson et al (2008) are analyzing the IMF 
for low mass, free-floating, objects whereas our concern is the IMF for 
brown dwarf companions to stars.  Burgasser et al. (2007) consider the 
mass and mass distribution of companions to late-F to K-type dwarfs within 
25 pc of Earth.  Their Fig. 1 suggests that, the wider the binary, the 
closer the companion mass function approaches the canonical field 
distribution.  Thus, the companion IMF depicted in our Fig. 2, being based 
on wide separation binaries, might mimic the field IMF.  Because almost 
all of the secondaries in young binaries upon which Fig. 2 is based have 
masses below the low mass cutoff of the Allen et al. and Anderson et al. 
studies, a direct comparison of our IMF with theirs' is not possible.

Concerning only companions (rather than field objects), it is of interest to
know how the brown dwarf companion mass function matches onto the very low mass
stellar companion mass function. The latter is a complicated issue (Burgasser
et al 2007; I.  N. Reid 2008, personal communication), and, to the best of our
knowledge, there is no published companion mass function that encompasses
secondary masses that straddle the stellar/substellar boundary. For example,
notwithstanding their interest in low mass companions, in the second edition of
their book, Reid \& Hawley (2005) declined to address the shape of the
companion mass function near the stellar/substellar boundary. 

In a paper that slightly postdated the Reid/Hawley book, Farihi et al (2005)
derived the companion mass function across the stellar/substellar boundary for
white dwarf primaries, i.e. stars that, when on the main sequence, were on
average more massive than the Sun.  Farihi et al found that late M-type (i.e.,
minimum mass stellar companion) are uncommon compared to mid M-type companions
(see their Fig.  6). Their survey of 261 white dwarfs capable of detecting
companions at orbital separations between $\sim$100 and 5000\,AU with masses as
low as 50\Mj\ (corresponding to the rightmost mass bin of our Figure 2) found
no brown dwarf companion.  Therefore, with the caveats that conclusions based
on young companions plotted in Fig. 2 suffer from small number statistics, and
the Farihi et al sample is limited to medium mass primaries, it appears that
companions with masses just above or below the brown dwarf/stellar dividing
line are rare indeed.  Recent model simulations of the formation of brown
dwarfs and very low mass stars are consistent with this conclusion (Stamatellos
\& Whitworth 2008).  Figures 5 and 13 of Stamatellos \& Whiteworth (2008)
illustrate that both very low mass stars and brown dwarfs, at the few hundred
AU semimajor axes of interest in the present paper, are expected to be uncommon
as secondaries to solar mass stars.

\begin{figure}
\begin{center}
  \includegraphics[width=0.95\columnwidth]{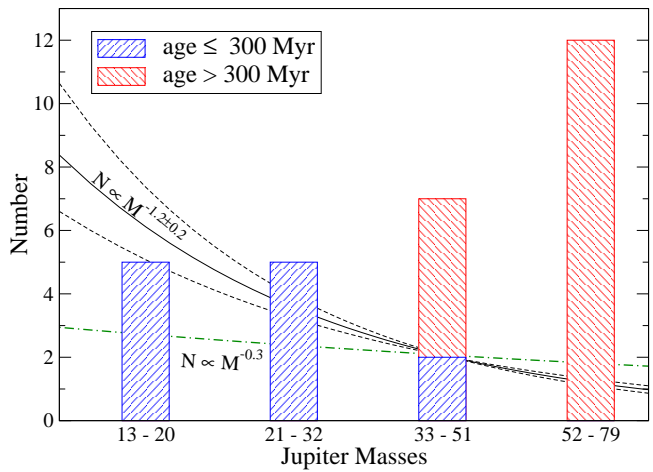}
\end{center}
\caption{Histogram of masses of secondaries with stellar primaries. 
Data are from Tables~1 \& 2 of this paper. For stellar primaries with 
brown dwarf secondaries that are themselves compact binaries (e.g., 
GL\,337B, G\,124-62B, etc.), we treat these compact binaries as singles 
by plotting total masses. The brown dwarf mass distribution is 
certainly flatter than that of Salpeter ($N\propto M^{-2.35}$).  
Considering only systems with ages $\leq$300 Myr (see Section 2.3), 
the number of secondaries is proportional to about M$^{-1.2}$. 
As described in the text and displayed in Table 1, there is a sharp 
secondary mass cutoff near 15 Jupiter masses -- that is, for stellar 
primaries, no imaged secondaries with masses below 13 Jupiter masses 
are known.  The dot-dashed line indicates the brown dwarf mass distribution 
expected if the result for free-floating objects (N proportional 
to M$^{-0.3\pm0.6}$) derived by Allen et al (2005) for masses 
$>$40\,\Mj\ obtains all the way down to 13\,\Mj .}
\end{figure}

\section{Conclusions}

We have gathered from the literature those binary systems with imaged
companions of the least mass (Table~1 and Figure~2).  Given the very
large number of target stars and brown dwarfs observed in imaging
ground- and space-based searches for low mass brown dwarfs and high mass
planets, these Table~1 objects represent pretty slim pickings.  We show
that minimum Jeans mass fragmentation of an interstellar molecular
cloud, as described a long time ago (Low \& Lynden-Bell 1976), can
account for these data at least as well as any more recent model for the
production of brown dwarfs.  Similarly, gravitational instability in
massive protoplanetary disks (Rafikov 2005 and Stamatellos \& Whitworth
2008) might account for some/many of the observed systems.

To derive the IMF for brown dwarf secondaries to stellar primaries it 
is essential to consider only youngish systems because data presented 
in this paper show that telescope/detector sensitivities are 
often insufficient to detect old, low mass, brown dwarfs. We find that 
the number of brown dwarf companions is proportional to mass as 
M$^{-1.2\pm0.2}$ down to the bottom of the brown dwarf mass range, 
$\sim$13 Jupiter masses.  While this power law
index might not apply to free-floating field brown dwarfs, the
precipice in the companion mass function for masses below 13\,\Mj,
suggests that free floating objects with masses in the planetary range
will be rare.

The extreme rarity of imaged companions below $\sim$15 Jupiter masses
suggests that Y-type objects ($T_{eff}$$\lesssim$500\,K) will be imaged as
companions to very few, if any, stars with ages $<2$\,Gyr. Even for a star
system as old as 7\,Gyr, according to the models of \cite{Baraffe}, a
brown dwarf would have to be less massive than $\sim$25 Jupiter masses to
cool to 500\,K.  Imaging discovery of a Y-type companion to a substellar
primary is even less likely, at any age, given the absence of wide
companions to field brown dwarfs (e.g., Allen et al. 2007).  Thus, all in
all, Y-type secondaries should appear in imaging programs only
infrequently.  Given the hundreds of young stars surveyed in the planet
hunting programs listed in the Introduction and the number of low mass
brown dwarfs indicated in Fig. 2, for a well choosen set of old stars,
perhaps one in 100 might be orbited at large separations by a Y dwarf.

\begin{acknowledgements}
We thank B. Hansen, M. Jura, and Neill Reid for helpful suggestions and 
the referee for constructive comments.
This research was supported in part by a NASA grant to UCLA.
\end{acknowledgements}

\begin{table*}
\begin{minipage}[t]{\textwidth}
\begin{center}
\caption{Brown dwarf secondaries (M $\geq$25 \Mj) to stellar primaries.}
\begin{tabular}{cccccccl}
\hline \hline
Object & \mc2{Sp. Type} & Age & $M_{pri}$ & $M_{sec}$ & Sep.  & Ref. \\
\cline{2-3}
  & Primary & Secondary & (Myr) & (M$_{\odot}$) & (\Mj) & (AU) & \\
\hline
GJ 802          &   M5+M5 &   L6    &$\sim2000$   &  0.28* & 66  & 1.46 & Ireland et al. (2008) \\
SCR1845-6357    &   M8.5  &   T6    &$ 1800-3100 $&  0.1   & 45  &   4  & Kasper et al. (2007b)          \\
Gl 337          &   G8+K1 &   L8    &$ 600-3400  $&  1.74  &110* &  11  & Wilson et al. 2001            \\
G 124-62        &   dM4.5e&   L0.5  &$  500-800  $&  0.24  & 72* &  13  & Martin et al. (1999)            \\ 
Gl 779B         &   G1V   &   L4.5  &$ 1000-3000 $&  1.02  & 66  &  13  & Liu et al. (2002)              \\
2MASS J1707-05  &   M9    &   L3    &$  500-5000 $&  0.077 & 70  &  15  & McElwain \& Burgasser (2006)     \\
Gl 86           &   K1V   &   L/T   &$ 1000-9999 $&  0.77  & 50  &  19  & Els et al. (2001)       \\
G 239-25        &   M1.5  &   L0    &$           $&  0.32  & 75? &  30  & Golimowski et al. (2004)        \\
HD 49197        &   F5    &   L4    &$  260-790  $&  1.4   & 54  &  42  & Metchev \& Hillenbrand (2004)    \\
Gl 229          &   M1/2V &   T7    &$   3000    $&  0.56  & 35  &  45  & Nakajima et al. (1995)          \\
HD 130948       &   G2V   &   L4+L4 &$   <800    $&  1.00  &140* &  47  & Potter et al. (2002)            \\
GL 569 B        &   M2.5V &         &$  250-500  $&  0.50  &123* &  49  & Lane et al. (2001)                    \\
GJ 1001         &   M3.5  &   L5    &$  >1000    $&  0.4   &100* & 178  & Goldman et al. (1999)           \\
HR 7329         &   A0Vn  &   M7/8  &$    12     $&  2.90  & 30  & 198  & Lowrance et al. (2000) \\
GG Tau B        &   M5    &   M7    &$    1-2    $&  0.12  & 44  & 207  & White et al. (1999)    \\
LHS 5166        &   dMe4.5&   L4    &$  <2600    $&  0.24  & 70  & 228  & Seifahrt et al. (2005)  \\
GJ 1048         &   K3V   &   L1    &$   1000    $&  0.72  & 65  & 250  & Gizis et al. (2001)             \\
GSC 8047-232    &   K3V   &   L0    &$  10-50    $&  0.80  & 25  & 279  & Chauvin et al. (2005b)     \\
G 196-3         &   dM3Ve &   L2    &$    100    $&  0.25  & 25  & 300  & Rebolo et al. (1998)            \\
DH Tau          &   M0.5V &   L2    &$  0.1-4    $&  0.33  &  ?  & 330  & Itoh et al. (2005)     \\
ScoPMS214       &   K1IV  &   M6    &$     5     $&  1.02  & 25  & 450  & Metchev (2006)            \\
HD 3651         &   K0V   &   T7.5  &$ 1000-9999 $&  0.79  & 40  & 480  & Mugrauer et al. (2006)          \\
HD 203030       &   G8V   &   L7.5  &$  130-400  $&  1.60  & 23  & 487  & Metchev \& Hillenbrand (2006)    \\
GJ 618.1        &   M0V   &   L2.5  &$ 500-12000 $&  0.51  & 70  & 1089 & Wilson et al. (2001)            \\
$\epsilon$ Indi Ba+b   &   K4.5Ve&   T1+T6 &$ 800-2000  $&  0.77  & 75* & 1459 & Scholz et al. (2003)            \\
Gliese 570      &  K5+M1+ &   T7    &$ 2000-9999 $&  1.7*  & 50  & 1526 & Burgasser et al. (2000)        \\
Gl 417          &   G0V   &   L4.5  &$   80-300  $&  1.0   & 35  & 1955 & Kirkpatrick et al. (2000)       \\
HD 89744        &   F7V   &   L0    &$ 1500-3000 $&  1.48  & 78  & 2460 & Wilson et al. (2001)            \\
Gl 584          &  G2+G2  &   L8    &$ 1000-2500 $&  2.0*  & 60  & 3612 & Kirkpatrick et al. (2000)       \\
\hline
\end{tabular}
\end{center}
For multiple systems consisting of a close brown
dwarf binary to a stellar primary, total mass of the brown dwarf
binary is listed in the table as $M_{sec}$ with `*' mark.
Likewise, when the primary is a multiple system, the total mass of
stars is listed as $M_{pri}$.
\end{minipage}
\end{table*}
\end{document}